# Framework for Ubiquitous Social Networks

Atta ur Rehman Khan[1], Mazliza Othman[1], Abdul Nasir Khan[1], Imran Ali khan[2]
[1]Faculty of Science and Information Technology, University of Malaya, Kuala Lumpur, Malaysia
[2]Department of Computer Science, COMSATS Institute of Information Technology, Abbottabad, Pakistan
attaurrehman@siswa.um.edu.my, mazliza@um.edu.my, anasir@siswa.um.edu.my, imran@ciit.net.pk

*Abstract–* **This paper presents a novel framework for ubiquitous social networks (USNs). Instead of making virtual connections, on the basis of human social networks, an effort has been made to facilitate interactions among human social networks with the help of virtual social networks. The imperative domains that support ubiquitous social networks are highlighted and different scenarios are provided to project real world applications of proposed framework. Our proposed framework can provide preliminary foundations for creating "ubiquitous social networks" in true essence.**

## I. INTRODUCTION

In today's world, full of informative technologies, it is becoming really important for people to have considerate and diversified social connections. To establish such connections, it is good to know people from different social backgrounds, expertise domains, career levels, interests and goals. Opportunities of making such connections are usually available, but people are unaware of such connections or hesitant to network with unknown people. To nullify the affect of such scenarios, a framework has been proposed to support ubiquitous social networking in true essence. This framework is not only beneficial in providing opportunities for making social connections but also provides a foundation for realizing social networks in a totally different approach.

Ubiquitous social networks (USNs) depends on different technologies and only with the support of such technologies, the existence of USNs can be made possible. The three imperative domains required to support USNs are:

a) Social Networks

b) Wearable Computers

c) Mobile Ubiquitous Computing

### A. Social Networks

In recent years, social networks have shown enormous growth and are continuing to grow rapidly [1]. Cheap cell phones and internet packages have played a vital role in the growth of social networks. In addition, the usage of personal computer for access to social networks is no longer required [2]. Due to these reasons, the usage of smart phones for social networks is continuously increasing.

Social networks have received enormous appreciation in recent years and among them the top most visited social networks are Facebook, MySpace, Twitter and LinkedIn [3]. Facebook is mainly used for entertainment purposes and to connect people with their family and friends. LinkedIn, on the other hand, tends to be professional and business oriented, where people collaborate with professionals for business purposes [4].

Considering the scenarios in the paper, the functionality of this framework is not limited to highlighted applications. Our proposed framework can support all types of virtual social networks, as far as the social network providers allow the use of such frameworks (APIs support for features).

### B. Wearable Computers

Since the dawn of computer age, computers are dumb and sit on the desk. Different areas of computer science are under research but not a single person is able to answer a very basic question. How should the computers be used? Or what is the true way of using a computer? Now, with the advancements in pervasive computing and wearable computers, it seems that finally, scientists would be able to answer this challenging question.

People have always dreamed of ubiquitous wearable computers that can interact with the users intelligently and according to the context of the situation. Ubiquitous wearable computers are assumed to be smart and efficient that they can make our daily tasks easier. Such wearable computers could be worn on the body or became a part of the clothing. A good example of such wearable computers can be a badge, a pair of glasses, a hand band, head mounted display or data input glove. Some scientists do not accept the mentioned devices as wearable computers and provides their own definition. They believe that a wearable computer should act like a part of a body and must not be considered as a separate device [5].

One may ask a question about smart phones, "are they wearable computers?". The answer could be "Yes and No", depending on how you define wearable computers. Smart phone are now equipped with sophisticated features like Wi-Fi, GPS, sensors, cameras and high computational processors, which were not imaginable few years back. Hence, smart phone is considered a positive step towards true wearable computers.

*C. Mobile Ubiquitous Computing*

Mobile ubiquitous computing refers to context aware seamless interaction of devices and users with their surrounding environment while on the move, in order to facilitate users by achieving defined goals. This domain is emerging rapidly due to its promising exiting feature of computing anytime and anywhere. Ubiquitous computing has already shown positive results by supporting important applications of multiple domains that include medical, military, habitat, smart environments etc.

There are three main aspects of mobile computing, namely mobile hardware, mobile communication and mobile software [6]. Mobile computing differs from traditional computing in that it does not require traditional computers and networks. In addition, it is usually user location aware.

Mark Weiser [7], who coined the term ubiquitous computing for the first time, stated that "The most profound technologies are those that disappear". The motivation behind this statement was the vision that computers and computing will be embedded in everyday appliances and environment, which will create smarter environments. Today, we have smart homes that are capable of sensing human presence and setting up home environment according to desired conditions with minimal or no user interaction.

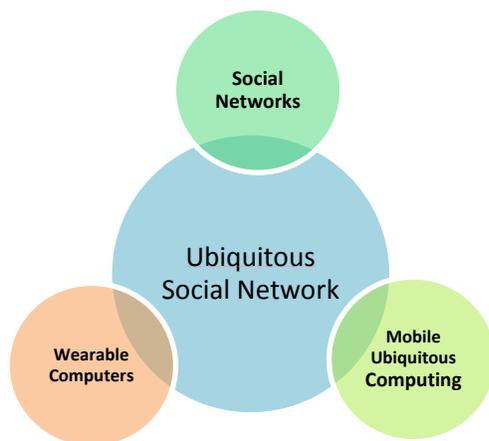

Figure 1: Ubiquitous Social Network

USN is basically a merger of the aforementioned technologies that have already achieved their promised objectives. Fig. 1 shows USN supported by the aforementioned domains.

The rest of this paper is organized as follows. In Section II we describe the related work. Section III explains proposed framework. In Section IV applications and scenarios are described and finally the paper is concluded in Section V.

## II. RELATED WORK

To the best of our knowledge, this type of framework has never been proposed. Therefore, we discuss the most prominent work that has been done regarding ubiquitous social networking below:

The iBand [8] is a wearable bracelet-like device that stores personal contact information. When two persons shake hands, the iBand recognizes the handshake gesture and exchanges the contact information between the two users. The users can later download the received contacts on a system named kiosk. To make iBand more attractive, it allows users to make their own personalized logo which shows on top of the iBand and is exchanged along with the contact information. Hence, different logos rotate on top of the iBand, which indicates multiple received contacts. The main drawback of iBand is the contact sharing control. If a person shakes hand with anyone, the contact information is immediately exchanged between the users even if they do not want to exchange their contact information. Considering cultural differences, it is not must for the people to shake hands when they meet.

The Smart-its Friends [9] is a project funded in part by the Commission of the European Union. When Smart-its enabled devices are shacked together and they form a connection in between. When users wearing the devices go out of transmission range of each other, a particular action is triggered, depending on the application. For example, informing a parent about a child who is getting far away from the parent.

Lovegety [10] is an egg-shaped device that comes in male and female versions. The male and female versions are produced in blue and pink colors respectively, and are small enough to fit in hands. The owner of a device can set the mode of lovegety according to their mood like simple chat, singing or dating. When two opposite gender lovegeties comes within range of 4.5 meters, the owners are informed, hence, providing them with an opportunity to interact.

CharmBadge [11] is a hardware device about the size of a business card. It is a computerized event badge that stores contact information of its wearer. The users feed their contact information in the charmbadge, and that information is exchanged with other users upon coming in range of each other. All contacts are stored in timely order (based on interaction time), which can later be downloaded from the badge at a charmbadge information access center.

## III. PROPOSED FRAMEWORK

The proposed framework is based on existing social network API's. To fully utilize the potential of USNs, some enhancements are proposed in social networks API's. The main components of our proposed framework are as follows:

### A. UbiServ

UbiServ is the server component of the framework that is connected to the internet. The main responsibilities of the UbiServ include tracking people, authenticating itself (UbiServ) with social network and fetching required data from the social network by using social networks APIs. In addition, UbiServ is responsible for user registration in USN service area, managing users profile records retrieved from social networks, privacy management and serving Ubiquitous Social Network Device (USND) requests.

### B. Ubiquitous Social Network Device

Ubiquitous Social Network Device (USND) is a client side hardware component of the proposed framework. USND contains a unique ID through which the user profile is identified on the social network. The main responsibilities of this component is to register the user with UbiServ, indentify other USND devices, request and receive data from UbiServ and display the data to the user.

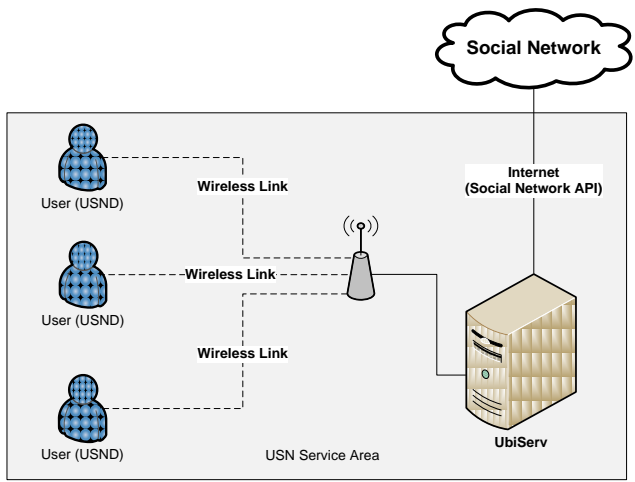

Figure 2: USN High Level Architecture

### C. Working

The initial prototype of USN is based on a distributed architecture and every USN service area consists of a separate server. USN service area is the social event coverage area that contains UbiServ and USND registration module. Whenever a user enters a USN service area, the UbiServ registers the user based on the USND unique ID. Once the registration is done, the USND device is authenticated to request profile from the UbiServ by providing the unique ID that is retrieved from USND of the target user. Upon receipt of the requested profile, UbiServ checks that both USND devices are within the same service area and the request is valid. Finally, the UbiServ requests the social network for a specific profile, which is served according to set permissions (on social network).

### D. Proposed enhancements in Social networks

We propose the following basic enhancements in the social network

- Social network users must be able to keep separate profile view for UbiServ requests in order to have good privacy control. For example, users may be interested in sharing their phone number at specific social events, instead of sharing it publically (online).

- The social networks APIs must be capable of recognizing and serving the UbiServ profile requests.

- Like normal user registration, UbiServ's must also be registered with social networks.

### E. Challenges

Currently, the main challenge in the implementation of this prototype is the ID retrieval procedure through which source USND will retrieve the ID of target the USND. Different technologies [12], [13], [14], are under consideration to overcome the aforementioned challenge that includes near-field communication, and RFIDs.

### F. Privacy concerns

As privacy always remains a primary concern when dealing with social network platforms, adequate efforts must be made to address privacy concerns. We propose the following considerations during the implementation of this system

- Users must be able to set permissions on their online profiles against UbiServ requests.
- Users must have control of disabling the USN service at USN Service Area (social event).
- UbiServ must make sure that the USND device is requesting data of only those users who are present at the event.
- UbiServ must be kept secure to avoid serving unauthenticated requests.

## IV. APPLICATIONS AND SCENARIOS

To highlight the importance and capabilities of USNs, different scenarios are provided as follows

### A. Conferences

There is a conference going on and different domain experts have gathered from all around the world to join that valuable event. During the tea break, all attendees gather in the gallery and starts chitchat with known people. A person is attending this conference for the first time and he does not know anyone. He would love to meet experts from his domain and make new connections, but unfortunately there is no one to help him.

He draws his USND and points it at a gentleman, standing alone. The device shows the person name, location, work domain and contact information. That person works in a domain that matches his, and he walks towards him, calling him by his name. After that, he introduces himself and starts discussing issues related to their domain. Later on, they add each other on virtual social network and hopes for a long term relationship.

*B. Job Fares*

There is a job fare taking place in a mega mall. All well known companies have set their stalls, and are looking for competent graduates. Students are submitting their CV's and companies officials are trying to make new connections with other companies for joint collaborations.

One of the companies CEO is also at that event. He is watching a guy who is explaining something to his friends. The CEO gets inspired with the expressions and presentation of the guy, and wants to check his background. He draws USND from his pocket and points it at the student. The device shows the student's name, qualifications, experience, job interest and contact information. The academic background of the student seems impressive and he considers appointing him. On the way back to office, the CEO instructs the HR manager to call that candidate for an interview.

*C. Parties*

There is a social event going on and people are coming all around the city to make new friends. Among the people, there is a girl who moved to the city a few days ago for her studies. She is all alone and she doesn't know anyone. As she has to stay there for few years, she wants to make new friends. She is conscious about making new friends and wants to meet people who are of her interest.

With the help of USND, she checks the profiles and pictures of the people present there. Finally, she selects a few people who are of her interest and willing to make new friends.

## V. CONCLUSION

In this paper, we proposed a novel framework that can facilitate its users in real life by using online social networks. This framework will make the researchers/service providers to think about social networks in a totally different perspective. In addition, it will provide opportunity for developing wide range of unique social network applications. The prototype of proposed model is currently under development. As a future work, we intend to propose and implement a USN framework based on mobile cloud computing in order to leverage full potential of proposed USNs.